\documentclass[12pt,a4paper,fleqn]{article}

\usepackage[textheight=23cm,textwidth=15cm]{geometry}
\usepackage{amsmath}
\usepackage[version=3]{mhchem}
\usepackage{graphicx}
\usepackage[american]{babel}
\usepackage{here}
\usepackage[articletitle=true, maxauthors=150]{achemso}
\usepackage{siunitx}

\DeclareSIUnit{\au}{a.u.}

\newcounter{rxnalign}
\newcounter{maineqno}

\newenvironment{rxnalign}
{
	\setcounter{maineqno}{\value{equation}}%
	\setcounter{equation}{\value{rxnalign}}
	\align
}
{\endalign\setcounter{rxnalign}{\value{equation}}\setcounter{equation}{\value{maineqno}}}

\begin{document}
	
\begin{center}

\LARGE{\bf{
		Pathfinder -- Navigating and Analyzing Chemical Reaction Networks with an Efficient Graph-Based Approach
}}

\large{
	Paul L. T\"urtscher\footnote{ORCID: 0000-0002-7021-5643} and
	Markus Reiher\footnote{Corresponding author; e-mail: markus.reiher@phys.chem.ethz.ch; ORCID: 0000-0002-9508-1565}
}\\[4ex]

Laboratory of Physical Chemistry, ETH Zurich, \\
Vladimir-Prelog-Weg 2, 8093 Zurich, Switzerland

November 07, 2022

\end{center}

\vspace{-0.5cm}
While the field of first-principles explorations into chemical reaction space has been continuously growing, the development of strategies for analyzing resulting chemical reaction networks (CRNs) is lagging behind.
A CRN consists of compounds linked by reactions.
Analyzing how these compounds are transformed into one another based on kinetic modeling is a nontrivial task.
Here, we present the graph-optimization-driven algorithm and program \textsc{Pathfinder} to allow for such an analysis of a CRN.
The CRN for this work has been obtained with our open-source \textsc{Chemoton} reaction network exploration software.
\textsc{Chemoton} probes reactive combinations of compounds for elementary steps and sorts them into reactions.
By encoding these reactions of the CRN as a graph consisting of compound and reaction vertices and adding information about activation barriers as well as required reagents to the edges of the graph yields a complete graph-theoretical representation of the CRN.
Since the probabilities of the formation of compounds depend on the starting conditions, the consumption of any compound during a reaction must be accounted for to reflect the availability of reagents.
To account for this, we introduce compound costs to reflect compound availability.
Simultaneously, the determined compound costs rank the compounds in the CRN in terms of their probability to be formed.
This ranking then allows us to probe easily accessible compounds in the CRN first for further explorations into yet unexplored terrain.
We first illustrate the working principle on an abstract small CRN.
Afterward, \textsc{Pathfinder} is demonstrated in the example of the disproportionation of iodine with water and the comproportionation of iodic acid and hydrogen iodide.
Both processes are analyzed within the same CRN which we construct with our autonomous first-principles CRN exploration software \textsc{Chemoton} [\textit{J. Chem. Theory Comput.} \textbf{2022}, \textit{18}, 5393] guided by \textsc{Pathfinder}.

\newpage

\section{Introduction}

Chemical reaction networks (CRNs) are molecular transformation webs with compounds connected by chemical reactions. 
Various computational strategies have been reported in the literature to explore chemical reaction space with quantum
chemical methods and map out such CRNs.\cite{Sameera2016,Maeda2021,Dewyer2017,Simm2019a,Unsleber2020,Steiner2022,Baiardi2022} 
These advanced exploration techniques make CRNs with increasing numbers of compounds and reactions accessible.
Alongside, the challenge of analyzing such large and interwoven networks is slowly surfacing.
Due to the high connectivity of a CRN, it is nontrivial to assess how specific compounds are formed.
This complexity is due to the fact that a compound of interest is likely formed through various reaction channels, each channel consisting of a sequence of reactions.
This multitude of options raises the question which channel or path is most frequently followed.\\
Microkinetic modeling\cite{Hoops2006a, Turanyi2014, Proppe2016, Goodwin2018, Proppe2019a} to approach this problem may not
be viable, if very many coupled ordinary differential equations that describe all concentration changes are to be
considered and span largely different time scales.
Then, the time required to simulate concentration fluxes through such a CRN can be exceedingly long.
Moreover, the sequence of reactions that form a compound in question is not directly retrieved.\\
However, the CRN can be represented as a graph (see, e.g., Refs.\ \citenum{Habershon2016, Feinberg2019, Unsleber2020}). 
A graph is an obvious representation as all compounds are interconnected via reactions in a CRN.
For instance, a compound can be a reactant or reagent of one reaction and the product of another reaction.
While a reactant is consumed during the course of a reaction\cite{ChemistryIUPAC1997}, a reagent, for instance, a solvent molecule, does not have to be consumed. 
In a graph, reactant and reagent have to be handled identically.
Once the graph is established, it can be analyzed to identify the shortest simple path between two compounds.
Or, in other words, the best channel to form one compound from another.
A shortest simple path is of minimal length with the lowest possible sum of the edge weights and vertices that are only visited once.
The shortest path is to be understood as the most efficient path, not necessarily the path with the least number of steps (reactions).
In the context of this work, the shortest path is the most probable sequence of reactions from a source vertex to a target vertex expressed by the lowest possible sum of edge weights.
The shortest path is a simple path as visiting a compound twice would contradict the chemical question of how one compound is formed from another through a sequence of reactions.\\
Determining the most probable sequence of reactions between two compounds is a shortest path problem.
Extracting such paths in terms of a likelihood for reaction mechanisms and synthesis routes has been accomplished based on depth-first algorithms.\cite{Kowalik2012,Robertson2020,Grzybowski2022}
A drawback of these approaches is that they are specifically tailored solutions for a specially constructed and pruned CRN.
For example, the stoichiometric requirements of a reaction are not considered or circumvented by construction.
This makes these approaches not generally applicable and insufficient for larger networks.\\
Persson and co-workers employed efficient algorithms from graph theory to identify paths in a CRN.\cite{Blau2021,Xie2021}
This allowed them to find one or multiple shortest paths with Dijkstra's\cite{Dijkstra1959} and Yen's\cite{Yen1971} algorithms in CRNs containing only three types of reactions and thermodynamic information.
These authors extended their approach to problems of solid-state material synthesis
where the CRN consists of phases instead of compounds.\cite{McDermott2021}\\
For larger CRN, one could think of employing an advanced path searching algorithm like the A* algorithm.\cite{Hart1968, Hart1972}
However, this would require a heuristic to determine the distance between any two compounds.\\
For even larger CRNs, a graph-based approach to identifying connections between compounds is computationally too demanding and requires a Monte Carlo-based strategy.\cite{Barter2022}
Such an approach has been exploited to study the formation of solid-electrode interfaces.\cite{Spotte-Smith2022}
However, the challenge of analyzing a CRN is rooted in its construction.
The construction of the CRN in Ref.~\citenum{Barter2022} relies on the enumeration and filtering of all stoichiometrically valid reactions based on a set of compounds.
The compounds of a set have the same composition in terms of their atoms (identical molecular formula).
Reactions are then generated by choosing all possible combinations of two compounds in each set, each compound for one side of the reaction.
During the construction of a CRN, it is not validated whether the two sides of a reaction are actually connected through a transition state (TS).
Each set consisting of $m$ compounds results initially in $m!/(m-2)!$ reactions.
For instance, a set with 10 compounds yields 90 reactions and a set with 50 compounds yields already 2,450 reactions.
By applying filters to the initial reactions, the number of reactions decreases.
Still, the underlying combinatorial approach in Ref.\citenum{Barter2022} produced a CRN with about 86,000,000 reactions.\\
By contrast, the exploration strategy developed in our group \cite{Bergeler2015,Simm2017,Unsleber2022a} 
explores CRNs through \textit{ab initio} calculations.
Accordingly, all reactions discovered connect compounds at least via one elementary step with a transition state. Hence,
a combinatorial approach, which assumes that all reactions are possible followed by subsequent filtering, is avoided.\\
In this work, we present a unique graph-based approach to
represent any multireactant reaction in a CRN including kinetic and stoichiometric information that then
allows for the identification of shortest paths between any two compounds.
An overview of the reaction to graph conversion is depicted in Fig.~\ref{fig:graphStructure} in Section~\ref{sec:graphRepresentation}.
The term reaction is to be understood in the most general sense in this context; it
can involve catalysts, solvents, and surfaces.
This comprises, for instance, organocatalyzed Mannich reactions, Michael additions, Suzuki cross-coupling, or \ce{CO2} reduction on different metal oxide surfaces.
Building the graph is neither affected by the type of reactant nor does it require a tailored template to represent it.
The stoichiometric requirements encoded in the edges of the graph carry information about which reagents are consumed when 
one compound vertex is transformed into another one.
Including the stoichiometry alongside kinetic information for a specific reaction in the graph is crucial for the assessment of 
how costly it is to traverse the network through the corresponding reaction vertex in the graph, 
since it is, in addition to the reaction barrier, important to know whether the necessary reagents are available at all.
Then, the graph representation allows us to query the network for the shortest simple path between any two vertices and nodes, respectively, employing efficient pathfinding algorithms.
The analysis of a CRN based on given starting conditions returns a ranking of its compounds corresponding to the probabilities of encountering them.
Based on such a ranking, a running CRN exploration can be steered \textit{on-the-fly} and, therefore, focused on kinetically relevant areas of the network.
This overall approach could, for instance, help to explore synthetically expensive (in terms of time, chemical supplies, etc.) or hazardous reactions \textit{in-silico} to determine, if a targeted product is formed at all, what kind of (hazardous) side products might emerge during the course of a reaction and what is their ratio compared to the targeted product without relying on kinetic modeling (cf. Ref.\ \citenum{Toniato2022}).\\
We introduce our \textsc{Pathfinder} algorithm in Sec.~\ref{sec:theory}, apply it to an abstract network taken from the literature\cite{Blau2021} in Sec.~\ref{sec:abstractExample}, and illustrate its predictive power for the disproportionation of \ce{I2} with \ce{H2O}\cite{Dushman1904, Murray1925, Sebok-Nagy2004, Truesdale2003b}, 
\begin{align*}
	\ce{3I2 + 3H2O <=> HIO3 + 5HI},
\end{align*}
for which we provide a new CRN of gas-phase reactions.
In this case, \textsc{Pathfinder} fulfills two tasks: guiding the exploration and analyzing the final CRN to allow for a comparison of paths from \ce{I2} to iodic acid, \ce{HIO3}, and vice versa.
The chosen reaction is just one example to showcase the working principle of \textsc{Pathfinder}. 
\textsc{Pathfinder} can, in principle, be applied to any type of (emerging) reaction network.

\section{Theory}\label{sec:theory}

We adopt the notation for first-principles CRNs from Ref.~\citenum{Unsleber2022a}:
A chemical \textit{structure} is given by a fixed atom type and number, fixed nuclear positions, a fixed number of electrons, and total spin.
It therefore represents a specific point on the Born-Oppenheimer potential energy surface.
By contrast, a \textit{compound} denotes a set of chemical structures with the same atom types and number, charge, spin, stereochemistry, and, most importantly, connectivity.
For instance, the boat and chair conformations of cyclohexane, clearly different structures, are considered the same compound.
A \textit{reaction} is defined as a collection of elementary steps that connect compounds, whereas an elementary step within a reaction connects structures of these compounds by different transition states (TSs).

\begin{figure}[ht]
	\centering
	\includegraphics[width=0.85\textwidth]{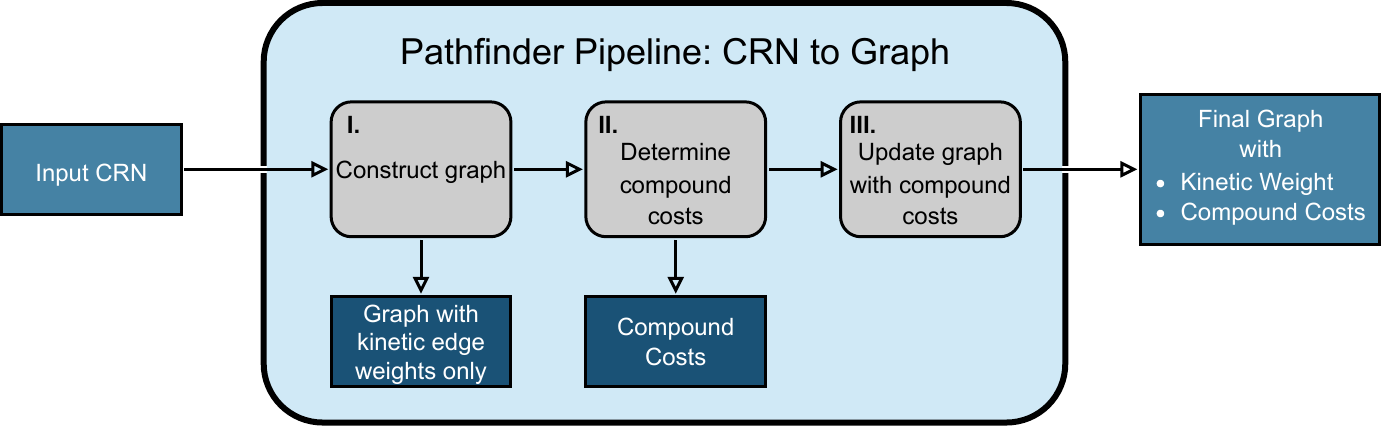}
	\caption{
		Schematic representation of the three steps necessary to convert a CRN to a graph containing kinetic as well as stoichiometric information.
		Input and output are colored in blue, operations sequentially performed on the input in gray, and the intermediate objects in dark blue.
		The kinetic information is added in step \textbf{I.} as edge weights, and the stoichiometric information in step \textbf{III.} by updating the edge weights with the compound costs.
	}
	\label{fig:pathfinderPipeline}
\end{figure}

Given a CRN defined for a set of compounds and reactions, one is interested in the kinetically most likely sequence of reactions within this CRN connecting two given compounds.
We dissect the search for such sequences into three steps:
First, a graph network is built from chemical reactions.
Second, a cost measure for the probability of a reactant being available for a reaction is evaluated.
The probability of a reactant being available can be understood as the likelihood of a reactant being spatially close enough for reacting with the original compound.
Thereby, the probability may be viewed as being related to the (local) concentration of the reactant and the compound of interest.
We refer to this cost as \textit{compound cost}.
We emphasize that ``costs'' are an abstract measure that convolutes various information necessary to provide a reasonable approximate picture for microkinetic modeling.
Third, the graph network is updated with the information about the costs of the reactants.
These three steps are sketched in Fig.~\ref{fig:pathfinderPipeline}.

The compound costs will depend on which compounds are available with some probability at the start.
Here, ``availability'' refers to reactants being present at the start of the reaction in the computer experiment, which may be interpreted as compounds being provided in a flask at the start of a reaction.
The probabilities can be derived from compound ratios of concentrations, pressure, or volume.
The starting conditions can be deliberately chosen.
Given the costs for at least one reactant, the costs for all other reactants follow (for details see below).

The resulting graph representation of the CRN can be queried for sequences of reactions, named paths in this context.
A typical query asks for the shortest paths starting from one starting compound to a compound of interest.
The obtained paths between two compounds 
are ordered by their path length which is linked to their probability.

	\subsection{Graph-Theoretical Representation of Reactions}\label{sec:graphRepresentation}
		
		In a given list of reactions (see Sec.~\ref{sec:compMethod} for details on their exploration), each reaction corresponds to a subgraph of the graph representation of the whole CRN.
		Therefore, the whole CRN is built by sequentially adding (parts) of such a subgraph.
		Two ``reaction vertices'' represent the TS structure of the reaction with the additional information from which side of the reaction the reactants approached the TS.
		Accordingly, they are labeled as ``left-hand-side'' (LHS) reaction vertex in case one starts from the reactants and ``right-hand-side'' (RHS) reaction vertex in case one starts from the products, although such an assignment is to a certain
degree arbitrary and solely serves the purpose to allow for directional distinction.
        In our context, the way a reaction is formulated depends on how our reactions exploration software \textsc{Chemoton} stored the reaction in the database.
        The assignment itself does not alter the graph as reaction vertices are only connected to reactants and products of a reaction. 
                		
        \begin{figure}[!h]
        	\centering
        	\includegraphics[width=0.9\textwidth]{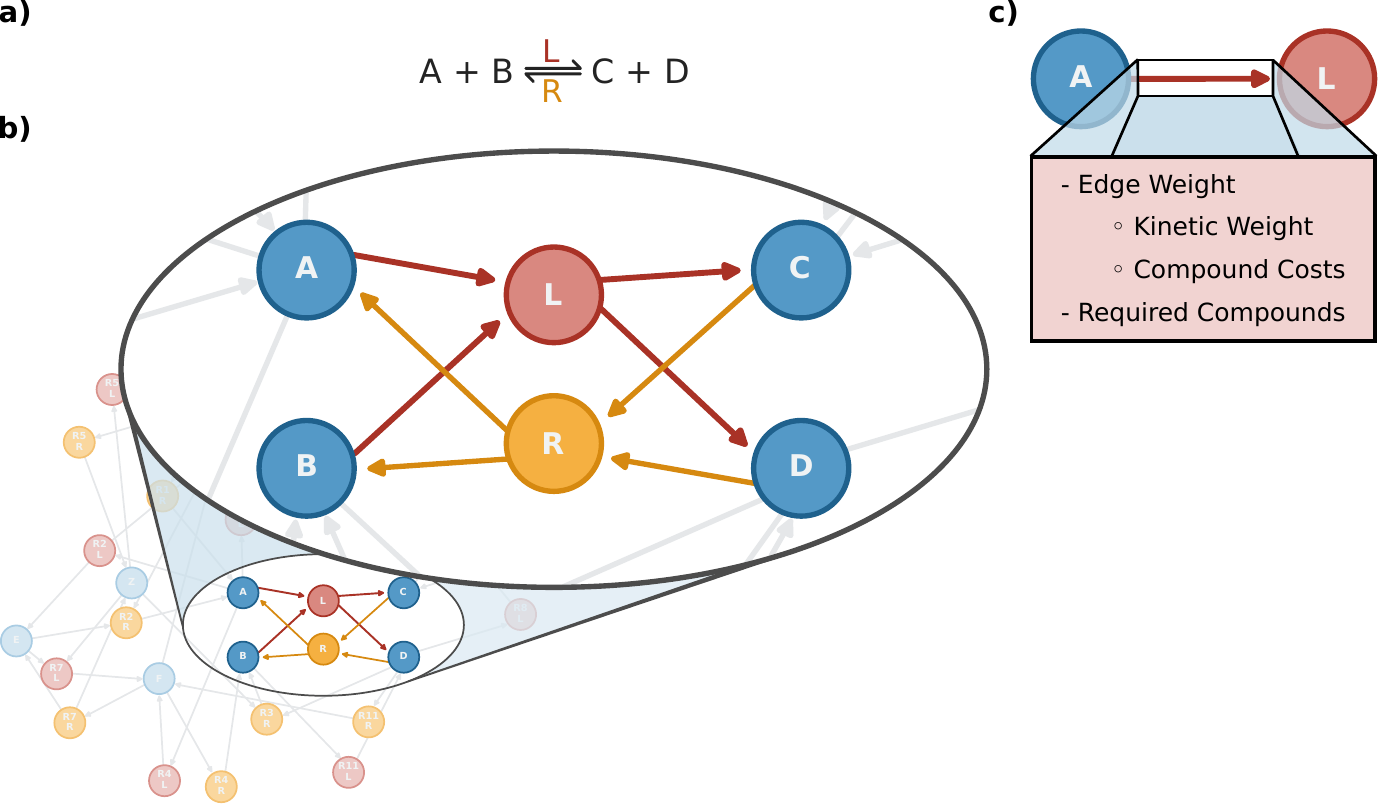}
        	\caption{
        		\textbf{a)} Prototypical chemical reaction in a CRN.
        		\textbf{b)} The same reaction in the CRN represented as a graph.
        		Compounds are depicted as blue vertices, the reaction vertex 'L' indicating the approach 
        		of the reactants toward the TS from the left hand side as a red vertex, and the reaction vertex 'R' indicating approaching the TS from the right hand side as an orange vertex. 
        		\textbf{c)} Properties stored in the edge from a compound vertex to a reaction vertex. 
        		Addends of the edge weight for the final graph, the kinetic weight and the compound costs.
        		In this case, the required compound in \textbf{c)} is \ce{B}.			     	    
        	}
        	\label{fig:graphStructure}
        \end{figure}

		Next, the reactants and products will be added to the graph as vertices if they have not already been included in the CRN yet.
		The added vertices are connected with directional edges where reactants are linked to the LHS reaction vertex and the LHS reaction vertex then to the products.
		In turn, the products are connected to the RHS reaction vertex and the RHS reaction vertex then to the reactants.
		The edges are directional to enforce the traversal from one side to the other as shown in Fig.~\ref{fig:graphStructure}.
		
		Furthermore, all edges contain an edge weight (measuring how costly it is to traverse this edge) and a list of required compounds.
		This list allows us to encode the information that, for instance, for a bimolecular reaction of the type \ce{A + A <=> B} two equivalents of \ce{A} are required without altering the directed graph structure or introducing parallel edges.
		Additionally, edges toward reaction vertices contain the sum over the compound costs of the reagents in the required compounds list.
		Strictly speaking, the resulting graph contains two types of edges.
		However, the difference is a mere technicality and has no effect on the performance of the presented algorithms.\\		
		All of these terms are explained in detail in Sec.~\ref{sec:infoEdge}.
		A schematic representation of this architecture is shown in Fig.~\ref{fig:graphStructure}.
		This architecture allows us to encode any kind of reaction from the CRN in the graph, regardless of the number of involved reactants or emerging products.\\

		The size of the graph measured in terms of the numbers of vertices and edges grows with each reaction added.
		At least two vertices will be added to the graph per reaction, if  all occurring compounds are already part of the graph.
		The maximum number of added vertices per reaction is given by two plus the number of compounds of the reaction which are yet to be included in the existing graph.
		Concerning the number of edges per reaction, the reaction vertices have incoming edges equal to the reacting compounds and outgoing edges equal to the number of product compounds.
		Hence, the number of edges per added reaction is the number of compounds of the reaction times two.
			
			The problem at hand is a shortest path problem; hence, a long edge indicates that the traversal via this edge is unfavorable. 
			The length of an edge, i.e., the edge weight, is added as information to the edge when connecting the vertices as shown in Fig.~\ref{fig:graphStructure}.
			An edge with a large edge weight would therefore correspond to a long edge in a graphical representation.
			Furthermore, the list of compounds required for this reaction is added.
			This list will be key for the second step of our algorithm to determine the costs for consuming compounds during a shortest path search.
			In the final step of the \textsc{Pathfinder} pipeline to convert a CRN to a graph containing kinetic as well as stoichiometric information, the sum over all compound costs of the compounds in the list of required compounds is written to the edge as compound costs.
			This sum is then added to the existing edge weight and allows us to consider the cost of consuming the required compounds of each reaction
			when querying the resulting graph directly.

		\subsection{Kinetic Weights}\label{sec:infoEdge}

			The initial weights for edges from compound vertices to reaction vertices, $\ce{C/D} \rightarrow \ce{R}$ and $\ce{A/B} \rightarrow \ce{L}$, are based on the free activation energies, $\Delta G_\text{TS-LHS}$ or $\Delta G_\text{TS-RHS}$, of the reaction.
			The free activation energies themselves are not suitable as edge weights as they are not reflecting the exponential character considering kinetics.
			Hence, the free activation energy is turned into a rate constant.
			The rate constant $k_i$, denoting $k_\text{LHS}$ and $k_\text{RHS}$ of reaction $i$ in a simplified notation, can be obtained according to Eyring's absolute rate theory,
			\begin{align}
			    \label{eq:Eyring}
				k_{i} = \frac{k_\text{B} T}{h} \text{exp}\left(- \frac{\Delta G_{\text{TS-}i}}{R T}\right),
			\end{align}
			 where $T$ is the temperature, $k_\text{B}$ is the Boltzmann constant, $h$ is Planck's constant, and $R$ is the ideal gas constant (accordingly, the energy difference is then to be given in energy units per mole substance).\\
			Note that the edge weights from reaction vertices to product vertices (e.g., $\ce{L} \rightarrow \ce{C/D}$ and $\ce{R} \rightarrow \ce{A/B}$ in Fig.~\ref{fig:graphStructure}) are set to zero throughout this work as the reaction progress from the TS to the product is energetically downhill in an elementary step. 
		
			 Since rate constants are not directly suitable as edge weights in a shortest path problem (owing to the anticorrelation that large rates would correspond to short paths), we consider a few additional steps.
			 Moreover, we note that rate constants can cover a large range of magnitudes, e.g., from \SI{1e-12}{s^{-1}} to \SI{1e+12}{s^{-1}}.
			 Hence, when adding edge weights, 
very small weights will hardly affect the total sum if a large edge weight is already part of the path.
A large range of weights is therefore disadvantageous for the rigorous assessment of the shortest path problem 
			 because paths going via different reactions besides one with a large edge weight would end up with the same length.\\
			 The correspondence of low reaction barrier to low edge weight as required for the shortest path problem
			 could be solved by taking the inverse of $k_i$, $1/k_i$ [\si{s}].
			 A low barrier would then correspond to a short time (low weight) to traverse an edge, but the problem of a large spread for the edge weights remains.
			 Hence, to minimize the spread we normalize each rate constant $k_i$ by the sum over all rate constants in the CRN to obtain the relative rate constant $p_i$ with
			 \begin{align}
			 	\label{eq:probability}
			 	p_i = \frac{k_i}{\sum\limits^{\text{all}\ k_\text{LHS}}_{l}k_l + \sum\limits^{\text{all}\ k_\text{RHS}}_{r}k_{r}}.
			 \end{align}
			 This definition can be rationalized by recalling that the rate $v_i$ of a reaction $i$ is given by
			 \begin{align}
				 v_{i} = k_{i} \prod^\text{reactants}_{j}c_{j},
			 \end{align}
			 where $c_j$ are the (time-dependent) concentrations of the reactants. If all
 concentrations of all reactants are considered equal to \SI{1}{mol/L} and time-independent, the rate $v_i$ will be
simply given by the rate constant $k_i$.
This assumption may also be understood literally as a short-time approximation to a case, in which all compounds
(also the stable intermediates usually produced only later in the course of a reaction starting from one or two reactants)
 are already available right from the start and at a concentration of \SI{1}{mol/L}. This assumption is necessary
to avoid explicit propagation of concentration flows through microkinetic modeling because this would be far too time-consuming. 
			 Under this short-time all-species-present assumption, the relative rate constant $p_i$ in a CRN is equal to the relative rate and can be interpreted as a measure for the likelihood or probability of the reaction step $i$ to occur relative to all other reaction steps in the CRN.
			 The probability of a sequence of reaction steps is the product of the individual reaction probabilities, assuming that each reaction is independent of all other reactions (cf. Eq.~(\ref{eq:probability})).
			 
			 Since a high probability of a path corresponds to a high likelihood that the path's product is formed through this reaction sequence,
			 it still contradicts the requirement of a low edge weight for a favorable shortest path.
			 Therefore, we introduce a cost function $f(p_{i})$,
			 \begin{align}
			 	\label{eq:costFunction}
			 	f(p_{i}) &=  - \text{ln}(p_{i}) = \text{ln}\left(\frac{1}{p_{i}}\right) \equiv w_{i},
			 \end{align}
                         which possesses the salient feature that it
			 results in weights $w_i$ which are additive to yield a total weight (instead of multiplicative probabilities that yield a total probability):
			\begin{equation}
			\begin{aligned}
				\label{eq:proofAdditivity}
				w_\text{tot} &= w_1 + w_2 + w_3 + \dots + w_i \\
				&= \text{ln}\left(\frac{1}{p_1}\right) +    \text{ln}\left(\frac{1}{p_2}\right) + 
				\text{ln}\left(\frac{1}{p_3}\right) + \dots +
				\text{ln}\left(\frac{1}{p_i}\right)\\
				&= \text{ln}\left(\frac{1}{p_1}\ \frac{1}{p_2}\ \frac{1}{p_3}\ 
				\dots\frac{1}{p_i}\right) = \text{ln}\left(\frac{1}{p_\text{tot}}\right),
			\end{aligned}
			\end{equation}		
			where $w_\text{tot}$ is the total weight or length of a path and $p_\text{tot}$ its total probability.
			 Note in this context that shortest-path algorithms typically sum over the weight of edges to determine a shortest path.\cite{NetworkX}
			The edge weights $w_i$ are given in arbitrary units abbreviated as \si{\au} throughout this work. 
			We call these weights derived from activation barriers entering our definition of the cost function \textit{kinetic weights}.
		
	\subsection{Compound Costs}
	
		If a compound can only be reached through reactions with high barriers, it may be unlikely that it can be formed under certain reaction conditions. 
		This fact must be considered when consuming the compound in another reaction, for which it will simply not be available.
		Hence, querying a CRN which encodes only the kinetic weights as edge weights would lack information about the stoichiometric requirement that compound C produced from compound A requires compound B as reactant in our example in Figure \ref{fig:graphStructure}.
		To take this into account, we introduce a \textit{compound cost} $c_i$ in arbitrary units (\si{\au}) for every compound $i$ in the graph.
		For one edge from a compound vertex to a reaction vertex, the additional costs caused by consuming the required compounds are encoded as the sum over the compound costs of all compounds required for a reaction.
		This sum of costs added to the edge weight $w_i$,
		\begin{align}
		\label{eq:updatedEdgeWeight}
			w_i^\prime &=  w_{i} + \sum\limits^{\substack{\text{required}\\ \text{compounds}}}_{l}\ c_{l},
		\end{align}
		results in $w_i^{\prime}$.
		$w_i^{\prime}$ of an edge $i$ is calculated on-the-fly when searching for the shortest path for determining the compound costs, the graph itself is not altered yet.
		A compound cost $c_i$ is defined as the shortest path from one of the starting compounds to compound $i$, considering the kinetic weight and the sum of compound costs of the required compounds,
		\begin{equation}
		\begin{aligned}
		\label{eq:compoundCostDefinition}
			c_i &= c_\text{start} + \sum\limits^{\substack{\rightarrow\text{R/L edges}\\ \text{in shortest path}}}_{j}\ w_j^\prime \\
			& = c_\text{start} + \sum\limits^{\substack{\rightarrow\text{R/L edges}\\ \text{in shortest path}}}_{j} \left( w_{j} + \sum\limits^{\substack{\text{required}\\ \text{compounds}}}_{l}\ c_{l} \right),
		\end{aligned}
		\end{equation}
		where $c_\text{start}$ corresponds to the compound cost of the path's starting compound.
		As the determination of the compound cost $c_i$ requires other compound costs, the costs must be determined iteratively, given the CRN and the starting conditions, to obtain self-consistent compound costs, $c_{i,\text{sc}}$.
		Hence, Eq.~\ref{eq:compoundCostDefinition} is rewritten as
		\begin{align}
		\label{eq:compoundCostDefinition2}
			c_{i,n} &= c_\text{start} + \sum\limits^{\substack{\rightarrow\text{R/L edges}\\ \text{in shortest path}}}_{j} \left( w_{j} + \sum\limits^{\substack{\text{required}\\ \text{compounds}}}_{l}\ c_{l,n-1} \right),
		\end{align}
		where $n$ indicates the current iteration step.
	
\begin{figure}[!ht]
			\centering
			\includegraphics[width=0.7\textwidth]{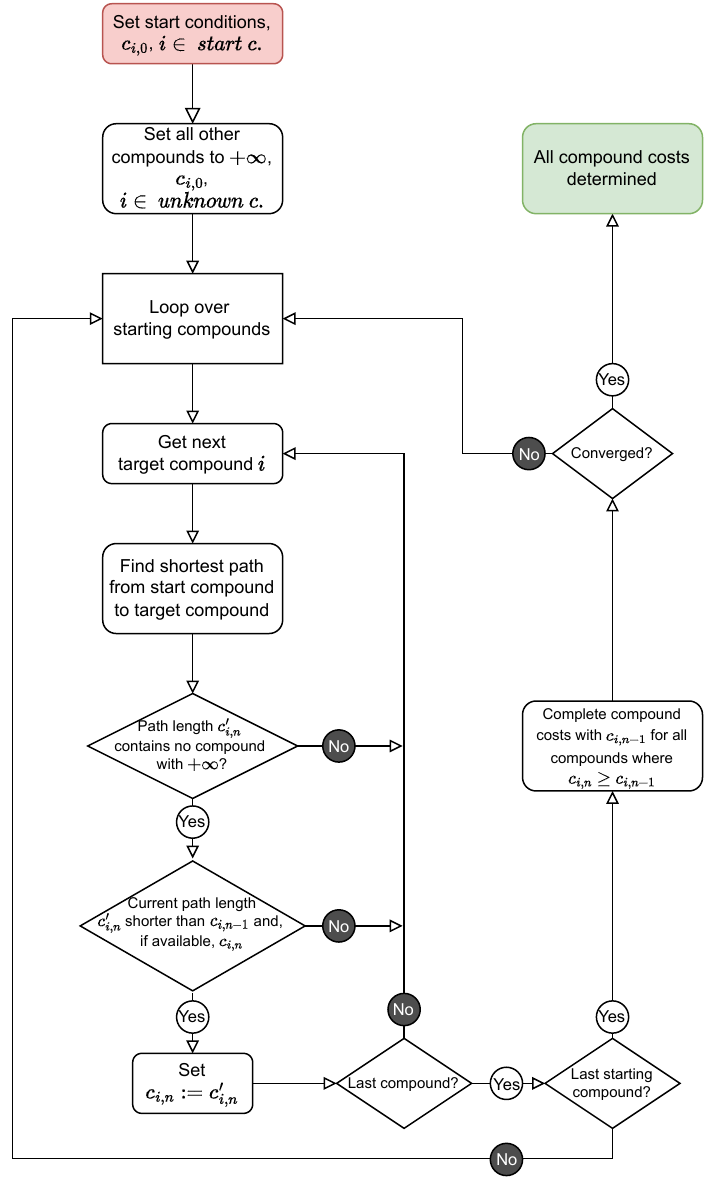}
			\caption{Flowchart of the algorithm to determine the compound costs in a CRN graph.}
			\label{fig:PRCost}
		\end{figure}

		The algorithm to determine the compound costs is outlined in Fig.~\ref{fig:PRCost}.
		To determine the required compound costs for all compounds in the CRN, we define starting conditions by assigning compound costs to selected starting compounds.		
		These compounds are those compounds available at the initialization of a reactive system under consideration, similar to reactants present in a flask at the beginning of a reaction in an experiment.
	                                                     
		When calculating costs of all other compounds, denoted here as ``unknown compounds'', the results will depend on the chosen starting compounds and their costs.		
		Hence, one can end up with different compound costs by choosing different starting conditions.
		Before starting the first iteration, all unknown compounds are assigned an infinite positive cost, 
		\begin{align}
		\label{eq:compoundCostsStart}
		c_{i,0} &=
			\begin{cases}
			0 \leq x < \infty,& i\in\text{starting compounds}\\
			\infty,& \text{otherwise}.
			\end{cases}
		\end{align}
		and with the compound costs for the starting compounds, all compounds have a $c_{i,0}$ in arbitrary units (\si{\au}) assigned.
		Then, the algorithm starts an outer loop over the set of starting compounds, the first iteration ($n=1$), and an inner loop over all unknown compounds.

		The shortest path from a starting compound to a target compound is found with Dijkstra's
		algorithm\cite{Dijkstra1959}.
		In principle, any other algorithm to determine the shortest path between the two compounds could be employed.\\
		If compounds of infinite costs would need to be consumed along the shortest path to a target compound, the algorithm stops any further analysis of this path and instead continues to find a path for the next target compound.
		If the shortest path does not require consumption of compounds with infinite cost, the compound cost of the starting compound is added to the current total path weight (see Eq.~\ref{eq:compoundCostDefinition2}) resulting in $c_{
		i,n}^{\prime}$ for the target compound $i$.
		The current total weight $c_{i,n}^{\prime}$ is calculated as stated in Eq.~\ref{eq:compoundCostDefinition2}.
		Employing compound costs of the previous iteration ensures that the order of compounds analyzed is not relevant during an iteration over all compounds.
		The current total weight $c_{i,n}^\prime$ for the target compound $i$ will be defined as $c_{i,n} := c_{i,n}^\prime$ if $c_{i,n}^\prime < c_{i,n-1}$ and, if $c_{i,n}$ has been previously defined, $c_{i,n}^\prime < c_{i,n}$.\\
		After iterating over all starting compounds, the compound costs of the current iteration are completed with $c_{i,n-1}$ for all compounds where the condition $c_{i,n}^\prime < c_{i,n-1}$ was not met.
		This guarantees that all compounds have a cost assigned for a possible subsequent iteration step.\\

		Finally, it is inquired whether the costs of the compounds have converged.
		Convergence will be achieved if all compounds have costs smaller than the assigned infinite value and self-consistency in the costs is reached.
		Self-consistency is achieved if none of the compound costs were altered in the current iteration.
		The loop over the starting compounds is restarted until convergence.
		Upon convergence, all compounds have costs assigned characterizing the length of the shortest paths from a starting compound to them and we achieve convergence, i.e.,
		\begin{align}
			c_{i,\text{sc}} := c_{i,n}
		\end{align}
		for all compounds $i, i\in\textit{unknown compounds}$.

	\subsection{Update Graph to Include Compound Costs in Edge Weights}
	
		With the self-consistent compound costs determined, the edges and edge weights of the graph are updated.
		The sum of all compound costs $C_i$ in edges $i$
		
		\begin{align}
			\label{eq:compoundCostsEdge}
			C_i = \sum\limits^{\substack{\text{required}\\ \text{compounds}}}_{l}\ c_{l,sc},
		\end{align}
		is stored in the corresponding compound-costs array.
		$C_i$ is then added to the edge weight $w_i$,
		\begin{align}
			\label{eq:finalEdgeWeight}
			w_{i}^{f} = w_i + C_i,
		\end{align} 
		resulting in $w_i^{f}$ which is set as the edge weight of an edge from a compound vertex to a reaction vertex (compare Fig.~\ref{fig:graphStructure}c).
		Consequently, the edge weight holds the kinetic information as well as the information about the accessibility of the required compounds to traverse along this edge.
		In this way, final edge weights in the updated graph of a CRN depend on the chosen starting conditions.
		By altering the starting conditions of the identical CRN, graphs with different compound costs and, consequently, different edge weights can be obtained and compared.\\
		The graph can now be queried from any source compound to any other target compound.
		The shortest paths between these two vertices are determined with Yen's algorithm\cite{Yen1971}, which in turn is based on Dijkstra's algorithm\cite{Dijkstra1959}.
		The returned path consists of a sequence of compound vertices and reaction vertices and the total weight or length of the path.			

\section{Computational Methodology}\label{sec:compMethod}

	\textsc{Pathfinder} implements our approach described so far in \textsc{Python3}. It can process network data produced by our autonomous first-principles CRN exploration software \textsc{Chemoton}\cite{Unsleber2022a,ScineChemoton}.
	The CRN's graph is represented and the shortest path is queried through the \textsc{Python3} package \textsc{NetworkX}\cite{NetworkX}.
	We note that the terms ``vertex'' and ``node'' are used synonymously in the source code following the notation in the \textsc{Python3} package \textsc{NetworkX}\cite{NetworkX}.
	\textsc{Pathfinder} is available open source and free of charge in \textsc{SCINE Chemoton}\cite{ScineChemotonV2.2} as well as in a slightly reduced form in \textsc{SCINE Heron}\cite{ScineHeron}.
	All data presented in this work were generated with this \textsc{SCINE} software framework\cite{scineFrame}, stored in and processed from the \textsc{SCINE} database.\cite{ScineDatabase}
	All calculations were handled by \textsc{SCINE Puffin} instances\cite{Unsleber2022a,ScinePuffin}.
	These instances interface \textsc{SCINE ReaDuct}\cite{Vaucher2018, Brunken2021}, \textsc{Molassembler}\cite{Sobez2020,Sobez2021}, and the \textsc{SCINE Utilities}\cite{Bosia2021}.
    New elementary steps were found by elementary-step trials with \textsc{SCINE Chemoton}\cite{ScineChemoton}.
    Details on the options for generating the trials are given in the Supporting Information (SI).
	To generate transition state guesses, the Newton Trajectory Algorithm~1 (NT1)\cite{Unsleber2022a} was selected.
	Detailed settings for all parts of the NT1 job are listed in the SI.
	\textsc{Chemoton} assigns newly found elementary steps to reactions as well as new structures to compounds in an automatic manner based on their definitions outlined in Sec.~\ref{sec:theory}, thereby constructing the CRN.
    We refer the interested reader to Ref.~\citenum{Unsleber2022a} for a detailed description of the underlying algorithms.
    
		For the construction of a chemical reaction network, a list of reactions is required.
		The elementary steps of each reaction 
	must consist of reactant structures 
	and assigned energies as well as product structures 
	and assigned energies. Both sets are then supplemented with (free) activation energies for both reaction directions.
		This information is easily accessible from an explored CRN when stored in a database.\cite{Unsleber2022a}
		In our framework, a reaction consists of multiple elementary steps, where each
		elementary step features its own barrier height. Hence, each reaction is assigned a range of activation energies,
                of which the lowest indicate those elementary steps with highest probability to dominate the reaction's mechanism.
		The obtained list of reactions can be subjected to additional filter criteria.
		For instance, only reactions with barriers in both directions below a given threshold may be forwarded to the ``CRN to Graph'' pipeline of \textsc{Pathfinder} (cf. Fig.\ref{fig:pathfinderPipeline}).
	The CRN generated with \textsc{Chemoton} here and the graph representation obtained with \textsc{Pathfinder} are available as the \textsc{IODAQ Exploration Data Set} on Zenodo.\cite{iodaq2022}
	To produce the raw data for the CRN, electronic structure Kohn-Sham density functional theory calculations were automatically launched by \textsc{Puffin} were carried out with the program package \textsc{Turbomole} (v7.4.1)\cite{TURBOMOLE}, applying the PBE functional\cite{Perdew1992, Perdew1996b} with a def2-SVPD basis set\cite{Peterson2003b, Weigend2005a, Rappoport2010a} and semiclassical D3 dispersion corrections\cite{Grimme2010}.
        All calculations were performed without imposing any point-group symmetry, i.e., in point group $C_1$.
	We note that almost all activation barriers were calculated as Gibbs free energy differences of transition state structures and stable intermediates, assuming the rigid-rotor/harmonic-oscillator/particle-in-a-box model at a temperature of \SI{298.15}{K} and a pressure of \SI{1}{atm} under ideal gas conditions. 
The Hessian matrix required for the vibrational analysis in the harmonic-oscillator model was calculated analytically with \textsc{Turbomole}.
	There are 17 exceptions where only electronic energy differences were available because of monoatomic molecules on either reactant or product site of the reaction.	
		For each reaction, only the free activation energies of the elementary step with the TS with the lowest free energy are considered as the key representative for a reaction under consideration.
		However, it is also possible to employ a (weighted) average over all (free) activation energies.

\section{Results and Discussion}

\subsection{Abstract Example}\label{sec:abstractExample}

		We first demonstrate the working principle of \textsc{Pathfinder} with the example of a CRN consisting of the following 10 abstract reactions:
		
		\begin{rxnalign}
			\ce{A <=>& B}\ &w_\text{R1} = 6\label{R1}\\
			\ce{A <=>& D + E} &w_\text{R2} = 1\label{R2}\\
			\ce{A <=>& Y + Z} &w_\text{R3} = 2\label{R3}\\
			\ce{A <=>& F} &w_\text{R4} = 4\label{R4}\\
			\ce{Z <=>& B} &w_\text{R5} = 3\label{R5}\\
			\ce{B + D <=>& K} &w_\text{R6} = 1\label{R6}\\
			\ce{E + Z <=>& F} &w_\text{R7} = 2\label{R7}\\
			\ce{Y + Y <=>& K} &w_\text{R8} = 4\label{R8}\\
			\ce{B + B <=>& H} &w_\text{R9} = 3\label{R9}\\
			\ce{F + K <=>& H} &w_\text{R10} = 1\label{R10}
		\end{rxnalign}					
	
\begin{figure}[!ht]
			\centering
			\includegraphics[width=0.95\textwidth]{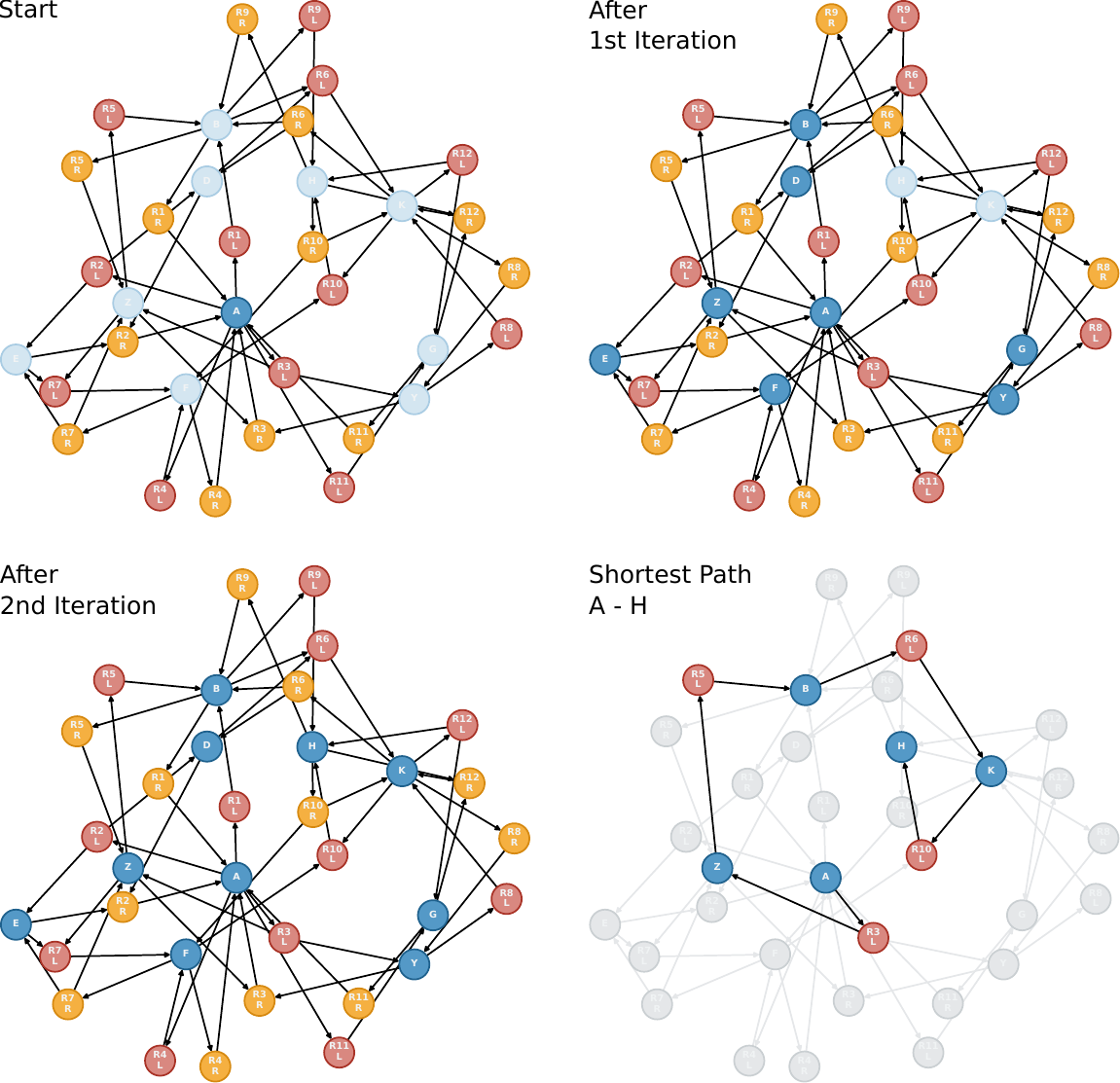}
			\caption{Determination of the compound costs in a model CRN of 10 reactions. Compounds with assigned costs are shown in dark blue, and compounds without assigned costs are in light blue. LHS reaction vertices are represented in red, and RHS reaction vertices are shown in orange. 
			At the start, only the starting compound has a compound cost (top left). After the first iteration step, all compounds, apart from compounds \ce{K} and \ce{H}, have a compound cost assigned (top right). After the second iteration step, also compounds \ce{K} and \ce{H} carry compound costs (bottom left). Not depicted here is the final third iteration step, which is performed to control if new paths including \ce{K} and/or \ce{H} can be found.
			Bottom right: the shortest path from compound \ce{A} to compound \ce{H} is highlighted, and all other vertices and edges are shown in gray.}
			\label{fig:compoundWeightOpt}
		\end{figure}

		The compounds occurring in the 10 abstract reactions are deliberately abbreviated by arbitrary letters.
		Reactions \ref{R1}--\ref{R10} and the initial edge weights in forward and backward directions were taken from Blau et al. \cite{Blau2021}
		Given only compound \ce{A} at the start of the reaction, we query the shortest path from \ce{A} to \ce{H}. At first glance, the
		shortest path in terms of the number of reactions is reaction~\ref{R1} yielding compound \ce{B} followed by reaction \ref{R9} producing compound \ce{H} directly from two equivalents of compound \ce{B}.
		However, the weight via reaction~\ref{R1} is quite high and hence a path along this reaction will be unlikely.
		The visual representation of the iterative determination of all compound costs (compare Fig.\ref{fig:PRCost}) is given in Fig.\ref{fig:compoundWeightOpt}.
		
		In the first iteration step of the \textsc{Pathfinder} algorithm, the compound costs (compare Fig.\ref{fig:compoundWeightOpt}) of products from reactions only directly consuming the starting compound \ce{A} as a reagent were assigned.
		Compounds \ce{H} and \ce{K} were not accessible in the first iteration as reactions forming those two require reactants not initially available.
		In the second iteration step, valid paths to the two remaining compounds were found such that all compounds ultimately had a cost assigned.
		In the third and last step, it was checked whether the availability of the two new compounds lowered any of the previously determined compound costs, which turned out to be not the case in this example.

		To probe the internal consistency of the compound cost determination, we added two reactions, reactions~\ref{R11} and \ref{R12}, to the CRN (reactions~\ref{R1}--\ref{R10}):
		
		\begin{rxnalign}
			\ce{A <=>& G} &w_\text{R11} = 8\label{R11}\\
			\ce{K + K <=>& G + H} &w_\text{R12} = 1\label{R12}
		\end{rxnalign}
		
		Although reaction~\ref{R11} is quite costly, the only path to \ce{G} was through this reaction in the first iteration step.
		As compounds \ce{L} and \ce{H} were only encountered in the second iteration step, the compound cost of \ce{G} could only be reduced when the costs for the other two compounds were determined.
		This was accomplished in the third iteration step which caused \textsc{Pathfinder} to iterate a fourth time.
		In the fourth step, it was probed whether the new compound cost of \ce{G} caused any other compound costs to change.

	\subsection{Exploration of \ce{I2} with \ce{H2O}}
	
		We now turn to a CRN produced with our \textsc{Chemoton} exploration software.
		The underlying raw data of the CRN was obtained from quantum mechanical calculations whose 
		details are given in Section~\ref{sec:compMethod}.
		The \textsc{Chemoton}-driven first-principles exploration was guided \textit{on-the-fly} by the ranking of compounds provided by \textsc{Pathfinder}.
		The initial starting reagents were chosen to be iodine and water only.
		The initial elementary-step trials were set up with only these two molecules, namely, combinations of either one iodine molecule and one water molecule, two iodine molecules, or two water molecules. In a setting, in which \textsc{Chemoton} reacts each emerging molecule with all existing molecules in the network, iodine compounds with more than one or two iodine atoms can be formed.
		\ce{I2} was assigned a compound cost of \SI{1.0}{\au}, \ce{H2O} a compound cost of \SI{0.45}{\au}.
		This corresponded approximately to a $\ce{I2}/\ce{H2O}$ ratio of $1:2$ and hence to a probability of \SI{36}{\percent} ($\text{e}^{-1.0}$) for \ce{I2} and \SI{64}{\percent} ($\text{e}^{-0.45}$) for \ce{H2O} at the start of the reaction.
		During the guided exploration, a total of 49,710~structures as well as 15,519~elementary steps from 100,997~elementary-step trials were found by \textsc{Chemoton}.

		The structures and elementary steps found were aggregated into a total of 1,157~compounds and 4,540~reactions.
Given the simplicity and small size of our starting molecules, this number of compounds is astonishing. However,
	as an iodine atom is able to form up to seven bonds, quite similar to transition metals, the first coordination sphere of iodine can vary significantly.
	For instance, orthoperiodic acid and diperiodate species coordinate six oxygen atoms per iodine atom.\cite{Wiberg2016}
Accordingly, the surprisingly large number of compounds is produced by a varying number of atoms bound to I and by their different types (and environments) -- similar to ligands in transition-metal complexes. In addition, stereoisomers (cis/trans, fac/mer) come into play.
As two iodine atoms can bind either directly or be bonded via a bridging oxygen atom, the number of theoretically possible compounds grows even more.
These compounds might be high in energy and, hence, meta-stable; nevertheless, they are minima on the investigated potential energy surface.
However, we emphasize that all species found in our network are, by construction, uncharged species in the gas phase. 
Hence, important solvation effects are absent, and therefore, charged species are not present.
		
		\begin{figure}[ht]
			\centering
			\includegraphics[width=0.85\textwidth]{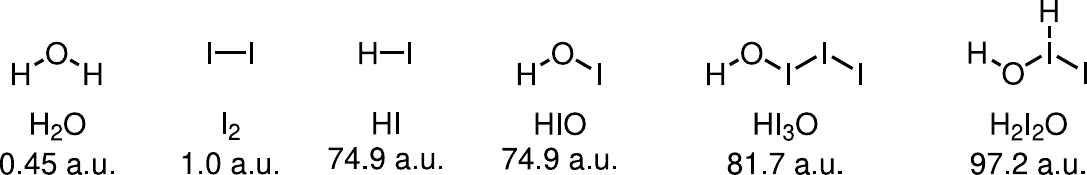}
			\caption{
				Lewis structures, molecular formulae, and compound costs determined with starting conditions set for the disproportionation of the six compounds with costs below \SI{100}{\au}.
			}
			\label{fig:Top6Compounds}
		\end{figure}

		During the exploration, the growing CRN was repeatedly analyzed with \textsc{Pathfinder} to determine the compound costs of all other compounds found.
		When starting from \ce{I2} and \ce{H2O}, all compounds with a compound cost below \SI{100}{\au} (six compounds in total) were allowed in unimolecular and bimolecular elementary-step trials in the CRN exploration process.
		Compounds with a compound cost $\SI{100}{\au} < c_{i, \text{sc}} < \SI{200}{\au}$ were only reacted with each other and with the six cheapest compounds, elementary-step trials with themselves were not permitted in the exploration process.
		For all compounds with costs above \SI{200}{\au}, only trials with the six cheapest compounds were probed to limit the total number of trials.\cite{Steiner2022}
		The rationale behind these choices is that the likeliest encounter of compounds with a high cost is with compounds of low cost.
		The six compounds with costs below \SI{100}{\au} are shown in Fig.~\ref{fig:Top6Compounds}, including the starting materials \ce{H2O} and \ce{I2}.
		All compounds with a cost below the one of \ce{HIO3} and including \ce{HIO3} were probed for trials in the exploration with \textsc{Chemoton}.

		After completion of these trials, 
the CRN was analyzed with different starting conditions
to investigate the reactivity of \ce{HIO3} with \ce{H2O} and the resulting products; namely, \ce{HIO3} and \ce{H2O} with compound costs set to \SI{1.0}{\au} and \SI{0.45}.
		\ce{H2O} had the same cost assigned as in the \textsc{Pathfinder} analyses before, the compound cost of \ce{HIO3} equaled the cost of \ce{I2} of the preceding exploration round.
		Due to the different starting conditions, compound costs were different compared to the analysis with \ce{I2} instead of \ce{HIO3} as starting conditions.
		Compounds with costs up to \SI{34}{\au} were only combined with \ce{H2O} in elementary-step trials as we were solely interested in reactions with \ce{H2O} at this point.
		
		The graph of the CRN was built from 3,916 of the 4,540 found reactions.
		A total of 624 reactions were not considered because either the forward or the backward reaction barriers were below \SI{4.5}{kJ / mol}.
		Reactions with barriers below this threshold were considered technical failures and therefore not included in the graph. 
		Hence, the resulting graph consisted of 1,046 compound vertices and 7,832 reaction vertices (LHS vertices and RHS vertices) with 29,212 edges.
		On a single core of an Intel Xeon E-2176G (3.70 GHz) central processing unit with a \textsc{Python} 3.8.13 interpreter, the construction of the graph took about \SI{2}{min} and the determination of all compound costs took about \SI{10}{min}.

		Among the discovered compounds were hydrogen iodide \ce{HI}, hypoiodous acid \ce{HIO}, iodous acid \ce{HIO2} and iodic acid \ce{HIO3}.
		These cover the oxidation states of iodine from \textrm{$-$I} to \textrm{V}.
		The disproportionation of \ce{I2} to \ce{HI} and \ce{HIO3} should therefore be included in the CRN as \ce{HI} and \ce{HIO3} were discovered starting from \ce{I2} and \ce{H2O}.
		
		\begin{rxnalign}
			\ce{3I2 + 3H2O <=>& HIO3 + 5HI} \label{R13}
		\end{rxnalign}
		
		The overall reaction equation of the disproportionation starting from \ce{I2} and \ce{H2O} is given in reaction~\ref{R13}.
		Mechanistic models for the formation of \ce{HIO3} have been discussed in the literature.
		Early attempts to describe the formation focused on the dissociation of iodine\cite{Murray1925} and proposed \ce{I2OH}, ``invented \textit{ad hoc}''\cite{Dushman1904}, as an intermediate to react with two equivalents of hypoiodous acid.
		In a more recent investigation, a kinetic model containing 10 proposed reactions was fitted successfully to the experimental observations.\cite{Sebok-Nagy2004}
		This model proposed \ce{I2OH-} as an intermediate to be crucial for fitting the experimental results.
		Iodate, \ce{IO3-}, was proposed to be formed from two equivalents of \ce{IO2-}.
		As a caveat, we emphasize that our exploration was performed in the gas phase (see \ref{sec:compMethod}).
		Hence, reactions leading to charged compounds, such as deprotonation reactions, are too high in energy in our setting that precludes dielectric stabilization effects and are therefore not observed in this work.

		We chose a kinetic model proposed in the literature\cite{Schmitz1987, Kolar-Anic1995} as our reference model and slightly modified it.
		The kinetic reference model for the disproportionation is part of a more complex kinetic model of the Bray--Liebhafsky reaction.\cite{Bray1931, Liebhafsky1931}
		The original kinetic model was postulated based on experimental observations and simulation attempts.
		Our kinetic reference model is represented in reactions~\ref{R14}-\ref{R17}.
		We neglected any dissociation reactions proposed in the original kinetic model in the literature due to the fact that solvent effects were not considered in our exploration.
		The proposed elementary steps for this kinetic reference model are then:
		
		\begin{rxnalign}
			\ce{I2 + H2O <=>& HI + HIO} \label{R14}\\
			\ce{2HIO <=>& I2O + H2O} \label{R15}\\
			\ce{I2O + H2O <=>& HIO2 + HI} \label{R16}\\
			\ce{HIO2 + HIO <=>& HIO3 + HI} \label{R17}
		\end{rxnalign}
		
\begin{figure}[!ht]
			\centering
			\includegraphics[width=0.95\textwidth]{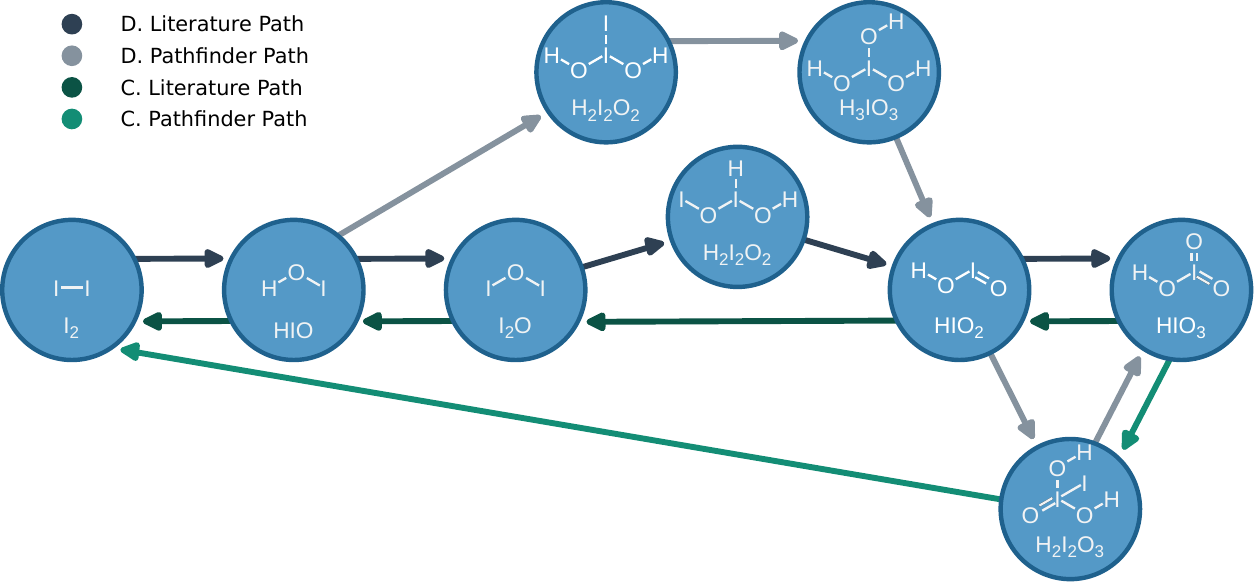}
			\caption{
				Summary of paths of the disproportionation of \ce{I2} and of the comproportionation of \ce{HIO3} and \ce{HI}.
				Compounds are represented as blue nodes with their respective Lewis structure and molecular formula.  
				Reagents and side products are not shown for clarity.
				Literature paths derived from experimental observations\cite{Schmitz1987,Kolar-Anic1995} for the disproportionation (D.) and the comproportionation (C.) are shown with dark gray and dark green arrows, respectively.
				The \textsc{Pathfinder} paths for the same start and target compound are shown with light gray and light green arrows.
			}
			\label{fig:resultsPath}
		\end{figure}

		To illustrate the operation of \textsc{Pathfinder}, we evaluated the literature path starting from \ce{I2} and \ce{H2O} (left-hand side of reaction~\ref{R14}) to \ce{HIO3} (right-hand side of reaction~\ref{R17}). 
		This corresponds to the disproportionation of \ce{I2}.
		Exchanging start and end of the path yields the comproportionation of \ce{HI} and \ce{HIO3}.
		We compare the literature path and its length to the shortest path found by \textsc{Pathfinder} for both directions, shown in Fig.~\ref{fig:resultsPath}, in the following sections.
		More detailed \textsc{Pathfinder} output on all paths is given in the SI.

		\subsubsection{Disproportionation of \ce{I2}}
		
\begin{figure}[!ht]
			\centering
			\includegraphics[width=0.95\textwidth]{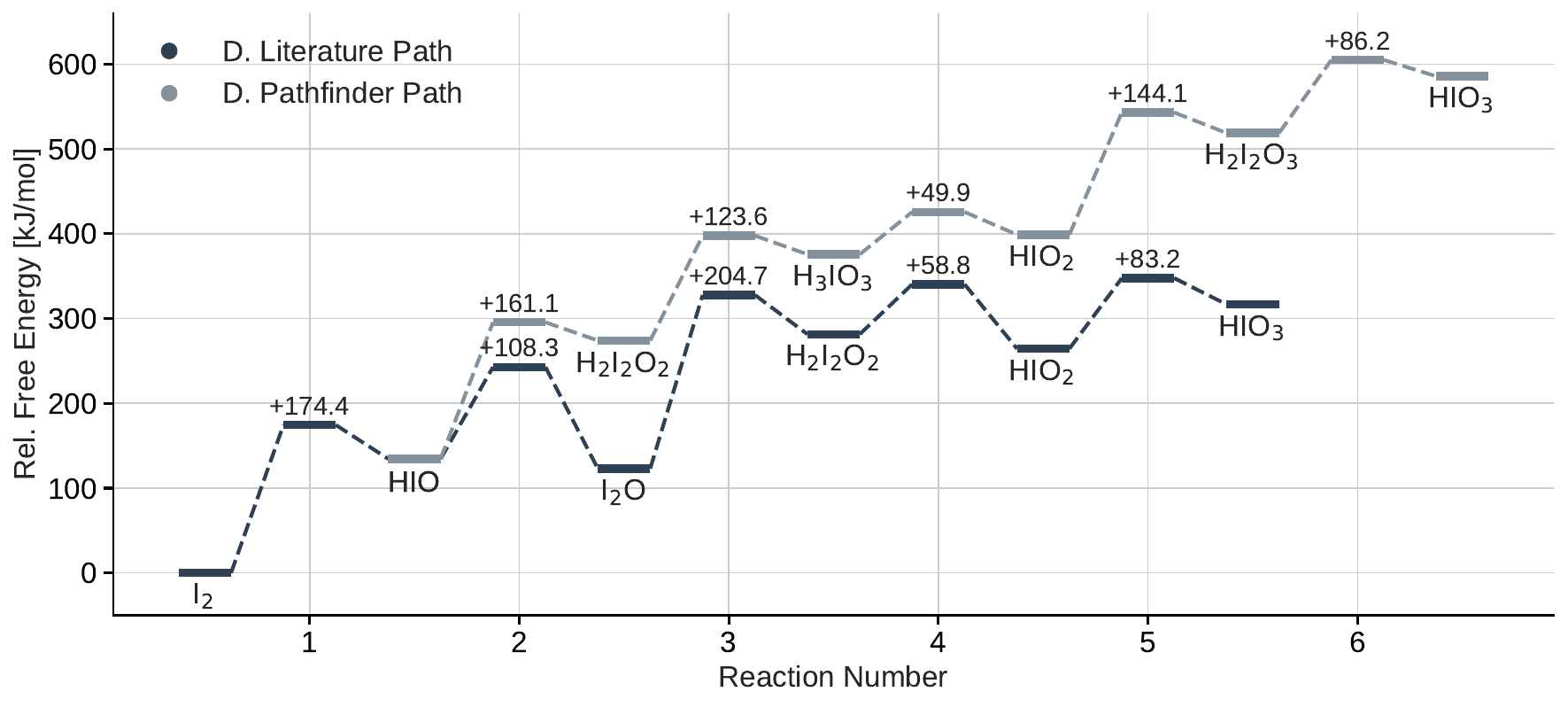}
			\caption{
				Free energies of the reactions traversed for the shortest disproportionation paths of \ce{I2} under consideration.
				For each reaction, free activation barriers are added (up) and subtracted (down), respectively.
				Free activation energies are given above the energy level of the TS of the corresponding reaction.
			}
			\label{fig:dispropReactionProfile}
		\end{figure}

\begin{figure}[!ht]
			\centering
			\includegraphics[width=0.95\textwidth]{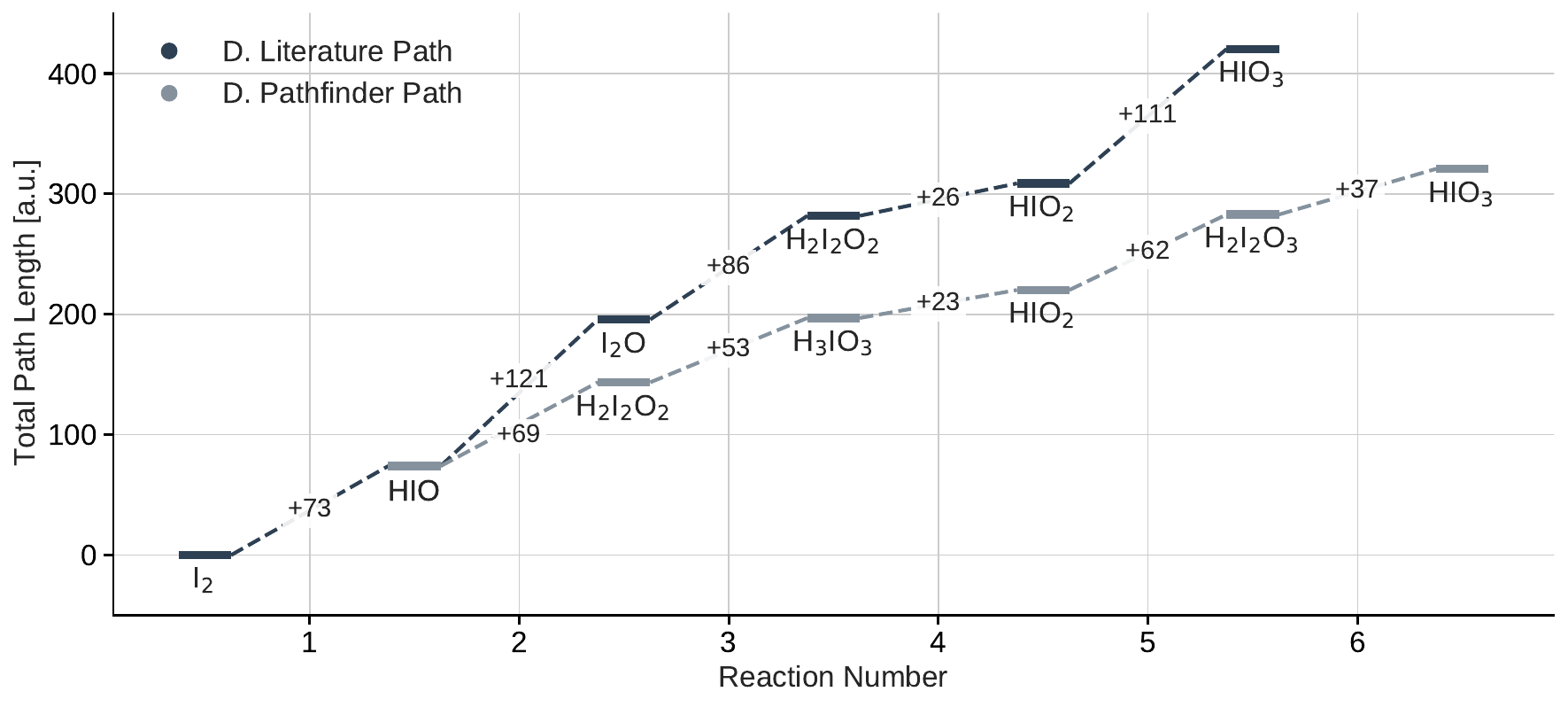}
			\caption{
				Total path length of the shortest disproportionation paths of \ce{I2} under consideration.
				For each reaction, the corresponding path length via this reaction is added and the length given.
			}
			\label{fig:dispropPathLengthProfile}
		\end{figure}

		An elementary step is defined as a chemical reaction in which the LHS and the RHS are connected by a single transition state.
		Reactions~\ref{R14}, \ref{R15}, and \ref{R17} were found in the CRN with elementary steps with one transition state (TS).		 
		However, reaction~\ref{R16} turned out to be not an elementary step in our CRN, as there was not one single TS connection between the left-hand side and right-hand side of reaction~\ref{R16}.
		Instead, the shortest path \textsc{Pathfinder} proposed to obtain \ce{HIO2} from \ce{I2O} under the given disproportionation conditions traversed via compound \ce{H2I2O2}, which has an \ce{HO-I-OH} motive (compare Fig.~\ref{fig:resultsPath}).
		The reactions traversed for the formation of compound \ce{HIO2} from \ce{I2O} included the highest free activation barrier along the literature path, as can be seen in Fig.~\ref{fig:dispropReactionProfile}.
		The added path lengths are shown in Fig.~\ref{fig:dispropPathLengthProfile}. 
		There were other possible paths with similar lengths.\\
		The total length of the reference path via the intermediate of \ce{H2I2O2} summed up to \SI{420}{\au}, as shown in Fig.~\ref{fig:dispropPathLengthProfile}. 
		Overall, two equivalents of \ce{HIO} were consumed along this reference path, namely, at reaction number two and five of the literature path.
		As the compound cost of \ce{HIO} was \SI{75}{\au} (compare Fig.~\ref{fig:Top6Compounds}) under the stated starting conditions, the total cost of the consumption of the two equivalents of \ce{HIO} summed up to \SI{150}{\au}.
		The consumption of two equivalents of \ce{HIO} therefore was responsible for \SI{36}{\percent} of the weight of the path's total length.

		The shortest path found by \textsc{Pathfinder} differed at the second reaction step from the reference path.
		Instead of forming \ce{I2O} and consuming one equivalent of \ce{HIO}, the path traversed via \ce{H2I2O2}, which has an \ce{IO-I-OH} motive (compare Fig.~\ref{fig:resultsPath}), and \ce{H3IO3} to form iodous acid \ce{HIO2}.
		The final steps to iodic acid differed as well, as it was more economical to react with equivalents of the starting compounds \ce{I2} and \ce{H2O} compared to \ce{HIO} in the reference path.
		This can be seen in Fig.~\ref{fig:dispropPathLengthProfile}, as reactions five and six of the path by \textsc{Pathfinder} elongate the total path by a length of \SI{99}{\au}, whereas the direct formation via reaction five of the literature path extends the total path by a length of \SI{111}{\au}.
		Hence, despite the reactions with higher activation barriers (compare Fig.~\ref{fig:dispropReactionProfile}), the path from \ce{HIO2} to \ce{HIO3} by \textsc{Pathfinder} was favorable in terms of path length because of the availability of the required compounds and consequently the lower compound costs.
		The overall total path length of the \textsc{Pathfinder} path was \SI{321}{\au} and thereby about \SI{100}{\au} lower than that for the literature path.
		The cost was reduced by avoiding the consumption of two equivalents of \ce{HIO} which were required in reactions~\ref{R15} and \ref{R17} of the literature path.
		
		The interpretation of a CRN by \textsc{Pathfinder} depends on the choices made for the exploration such as the underlying electronic structure theory, the modeling of the environment (gas phase vs solvation), and the starting conditions for determining the compound costs.
		Given the limitations of the performed exploration, especially the simplification to exclude any solvation effects, the difference in the reference and \textsc{Pathfinder} path are not surprising.
		The compounds along the \textsc{Pathfinder} path must also be considered with care as some have not been observed experimentally.
		However, the crucial point of considering the stoichiometric constraints for reactions is clearly covered by the \textsc{Pathfinder} analysis of a CRN as the reactions of the shortest path for the disproportionation are not consuming any reactants besides the starting materials \ce{I2} and \ce{H2O}.
		
		\subsubsection{Comproportionation of \ce{HIO3} and \ce{HI}}
		
\begin{figure}[!ht]
			\centering
			\includegraphics[width=0.65\textwidth]{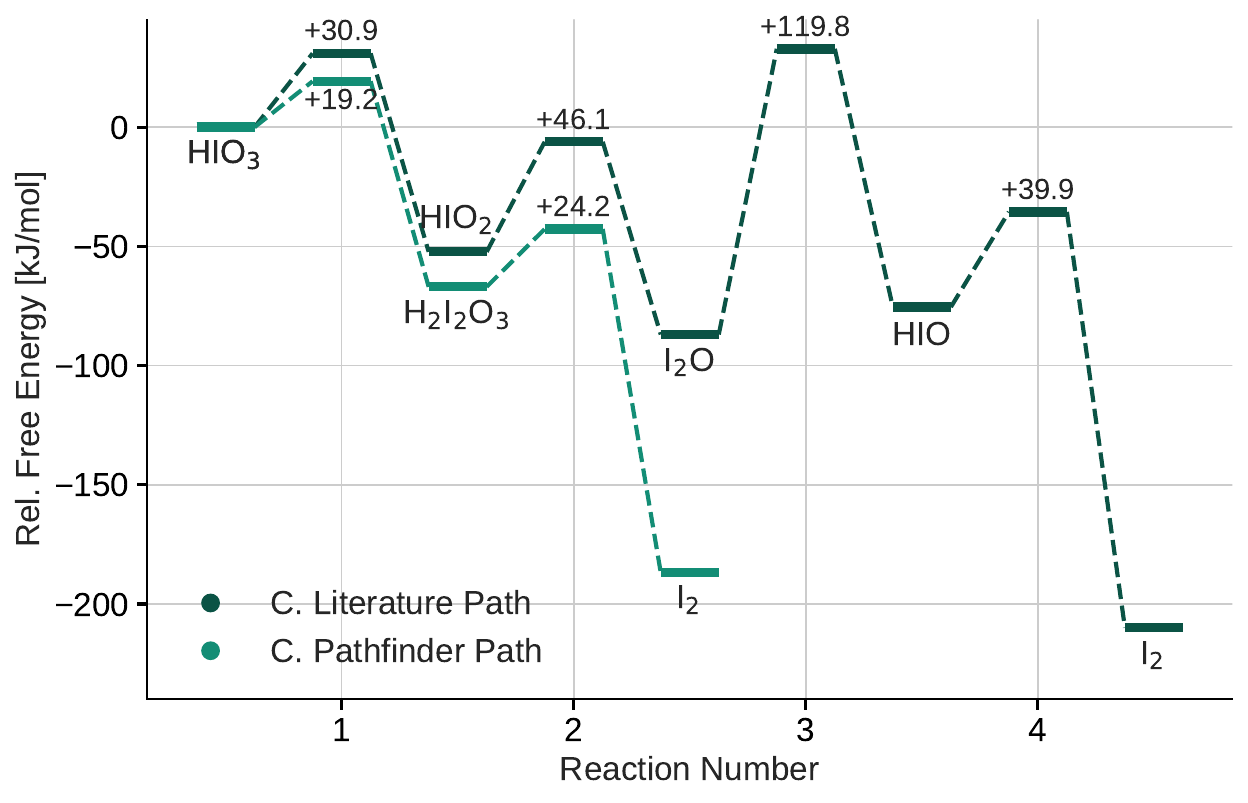}
			\caption{
				Free energies of the reactions traversed for the comproportionation paths of \ce{HIO3} and \ce{HI} under consideration.
				For each reaction, free activation barriers are added (up) and subtracted (down), respectively.
				Free activation energies are given above the energy level of the TS of the corresponding reaction.
			}
			\label{fig:compropReactionProfile}
		\end{figure}

		For the comproportionation, only \ce{HIO3} and \ce{HI} were considered as starting compounds.
		\ce{HIO3} was assigned a compound cost of \SI{1.0}{\au}, \ce{HI} a compound cost of \SI{0.45}{\au}.
		This corresponded to a probability of \SI{36}{\percent} ($\text{e}^{-1.0}$) for \ce{HIO3} and \SI{64}{\percent} ($\text{e}^{-0.45}$) for \ce{HI}, respectively, being available at the start.
		As we were analyzing the same CRN as for the disproportionation, the same reactions as discussed before were present.
		However, a path to form \ce{I2O} from \ce{HIO2} was still needed as reaction~\ref{R16} is not an elementary step in our network.
		Due to the different starting conditions, the shortest path to form \ce{HIO2} varied from the one found under the disproportionation conditions.
		\textsc{Pathfinder} identified a direct formation of \ce{I2O} from the reaction of \ce{HIO2} with two equivalents of \ce{HIO} as the shortest option.
		The steps found and discussed under disproportionation conditions for reaction~\ref{R16} were the second best option, only \SI{0.1}{\au} more expensive.\\
		With the starting conditions set for the comproportionation, the resulting compound cost for \ce{HIO} was \SI{17}{\au} instead of the \SI{74}{\au} determined under the disproportionation conditions.
		This illustrates the dependency of the determined compound costs on the chosen starting conditions.
		The easier formation of \ce{HIO} was due to the lower reaction barrier of \SI{30.9}{kJ / mol} from \ce{HIO3} to \ce{HIO2} (compare reaction~\ref{R17}), as shown in Fig.~\ref{fig:compropReactionProfile}, compared to the barrier of \SI{174.4}{kJ / mol} from \ce{I2} to \ce{HIO} (compare reaction~\ref{R14}).\\
		The highest reaction barrier in the literature path was the formation of \ce{HIO} from \ce{I2O}, reaction three of the path, corresponding to reaction~\ref{R15}.
		The overall cost of this reference path accumulated to \SI{159}{\au} with reaction three of this path elongating it by \SI{67}{\au}.
	
\begin{figure}[!ht]
			\centering
			\includegraphics[width=0.65\textwidth]{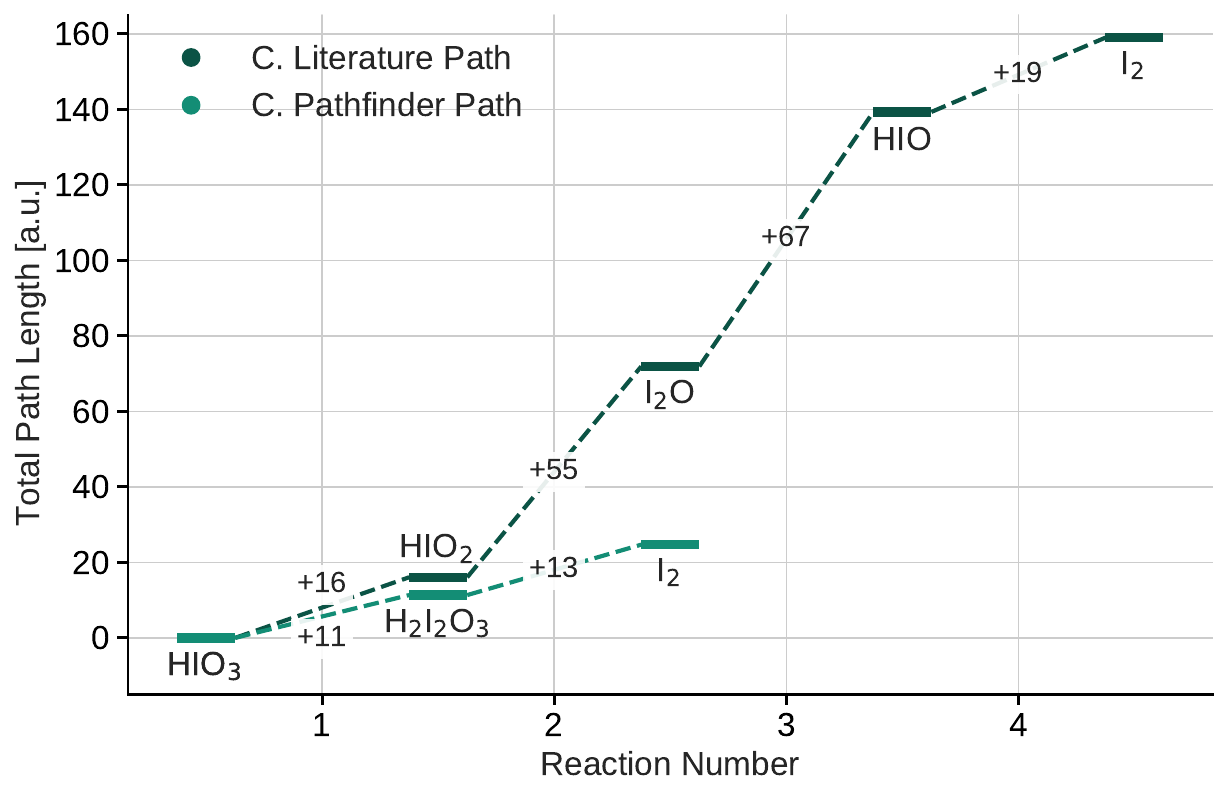}
			\caption{
				Total path length of the shortest comproportionation paths of \ce{HIO3} and \ce{HI} under consideration.
				For each reaction, the corresponding path length via this reaction is added and the length is stated.
			}
			\label{fig:compropPathLengthProfile}
		\end{figure}

		Finding the shortest path to iodine under the given starting conditions according to \textsc{Pathfinder} required only two reactions, as shown in Fig.~\ref{fig:compropReactionProfile} and Fig.~\ref{fig:compropPathLengthProfile}.
		First, compound \ce{H2I2O3} (compare Fig.~\ref{fig:resultsPath}) was formed from \ce{HIO3} and \ce{HI}.
		This intermediate then reacted again with one equivalent of \ce{HI}, a cheap starting compound, to form \ce{I2}.
		The overall length of this path was \SI{24.7}{\au}, more than \SI{100}{\au} lower than the literature path.
		The key feature distinguishing it from the reference path was that only starting compounds were consumed.
	
		As the reference model was based on experimental studies of the disproportionation,
         its validity for the comproportionation reaction has to be carefully considered.
		The fact that the shortest path according to \textsc{Pathfinder} was not overlapping with the literature path indicated that there might be other steps involved.
		\textsc{Pathfinder} allows for an almost unbiased approach, with the exception of the choice of the starting conditions.

\section{Conclusions}

	In this work, we introduced the \textsc{Pathfinder} algorithm for the fast and efficient analysis of general CRNs represented in a specific graph-based form. 
	In such graphs, kinetic weights are derived from reaction barriers obtained from quantum mechanical reference calculations.
	They are linked to the likelihood of a reaction in the CRN taking place.
	To consider the consumption of required reagents during a reaction, compound costs of all compounds in the CRN are determined under given starting conditions.
	When added to the kinetic weight, both the kinetic information and the reaction conditions of a reaction are encoded in the graph.
	This avoids explicit kinetic modeling of a CRN with many coupled ordinary differential equations causing exceedingly long simulation times.
        Instead, the shortest path is found by standard graph algorithms in a cost-efficient way.
	It is then directly possible to query how compounds are formed in terms of a reaction sequence, avoiding the post-processing of a kinetic simulation.

        We first demonstrated our algorithm with an abstract reaction mechanism.
	Then, we evaluated the disproportionation reaction of \ce{I2} with \ce{H2O}. 
	The CRN was generated with \textsc{Chemoton} and \textsc{Pathfinder}, where the latter \textit{on-the-fly} ranked the encountered compounds to guide the exploration by exploring only compounds with low costs.
	This strategy was followed until elementary-step trials of iodic acid with the six most probable compounds were probed. 
	The obtained CRN was further analyzed with \textsc{Pathfinder} and paths for disproportionation and comproportionation reactions were compared to paths proposed in the literature.
Despite the fact that our exploration only described gas-phase chemistry by construction, interesting observations could
be made regarding potentially important neutral species in these reactions.

	We emphasize that the graph architecture allows one to analyze the graph in other ways than those considered here.
	For example, running Monte Carlo simulations from a starting compound while at each compound vertex the next reaction vertex is chosen according to the compound vertices' relative out-flux could help to understand which compounds are formed.
Such a procedure would better approximate a full microkinetic modeling attempt on the network. Work along these lines is currently in progress in our laboratory.

\section*{Acknowledgments}
\label{sec:acknowledgments}
The authors gratefully acknowledge financial support through ETH grant ETH-44~20-1.
This work was presented at the WATOC 2022 conference in July 2022 in Vancouver.

\providecommand{\refin}[1]{\\ \textbf{Referenced in:} #1}

\end{document}